%% file: main.tex
\documentclass[runningheads]{llncs}

\usepackage{booktabs}
\usepackage{subcaption}
\usepackage{graphicx}
\usepackage{amsmath}
\usepackage{amsfonts}
\usepackage{comment}
\usepackage{hyperref}
\usepackage{xcolor}
\usepackage{siunitx}

\usepackage{booktabs}
\usepackage{adjustbox}

\newtheorem{assumption}{Assumption}[theorem]

\title{When Priority Fails:\\Revert-Based MEV on Fast-Finality Rollups}
\titlerunning{Revert-Based MEV on Fast-Finality Rollups}

\author{Krzysztof M. Gogol\inst{1} \and Manvir Schneider\inst{2} \and Claudio J. Tessone\inst{1,3}}
\authorrunning{K. Gogol, M. Schneider, C. Tessone}

\institute{University of Zurich UZH \and Cardano Foundation \and UZH Blockchain Center}

\begin{document}

\maketitle

\begin{abstract}
\input{sections/00Abstract}
\end{abstract}

\keywords{MEV \and Rollups \and Sequencer Design \and Priority Fee Auctions}

\input{sections/01Introduction}
\input{sections/02Background}
\input{sections/03Methodology}

\input{sections/05EmpiricalResults}

\input{sections/06Clustering}

\input{sections/07Theory-new}

\input{sections/08Discussion}
\input{sections/09Conclusions}

\section*{Acknowledgments} 
The first author gratefully acknowledges the support of the Flashbots Research Grant. The authors are also indebted to Danning Sui, Daniel Marzec, Johnnatan Messias, and Xin Wan for their valuable discussions and insights, which greatly contributed to shaping this research.

\bibliography{main}

\appendix

\input{sections/99SequncerOrdering}
\input{sections/98Appendix}
\input{sections/99Appendix}

\end{document}

%% file: sections/00Abstract.tex
We study the economics of transaction reverts on Ethereum rollups and show that they are not accidental failures but equilibrium outcomes of MEV strategies. Using execution traces from major L2s, we find that over 80\% of reverted transactions are swaps, with half targeting USDC–WETH pools on Uniswap v3/v4. 
Clustering reveals distinct bot archetypes—including split-trade arbitrageurs, atomic duplicators, and end-of-block spammers—demonstrating that reverts follow systematic patterns rather than random noise. Empirically, we show that priority fee auctions on rollups do not allocate blockspace efficiently: transaction placement is mis-ordered, round-number bidding dominates, and duplication spam inflates base fees. As a result, reverted transactions contribute disproportionately more to sequencer fee revenues than to gas consumption, shifting welfare from users to sequencers.
To explain these dynamics, we develop a model proving that trade-splitting and duplication strictly dominate one-shot execution under convex adversarial loss. Our findings establish reverts as a structural feature of rollup MEV microstructure and highlight the need for protocol-level reforms to sequencing, fee markets, and revert protection.

%% file: sections/01Introduction.tex
\section{Introduction}
    \label{sec:intro}


Maximum Extractable Value (MEV)~\cite{daian2020flashboys} has evolved significantly since the early days of Ethereum. The introduction of MEV Boost mechanisms, followed by Proposer-Builder Separation (PBS), transformed how MEV was captured on Ethereum, and now, with the rise of rollups~\cite{Thibault2022SoKRollup}, Layer-2 (L2) scaling solutions for Ethereum, history appears to be repeating itself.


The centralization of sequencers significantly impacts MEV on rollups. A sequencer is the entity responsible for ordering transactions, forming blocks, and submitting them to the Layer 1 chain~\cite{motepalli2023sokcentralizedsequencer}. As the sole actor capable of transaction ordering, the sequencer introduces implicit trust assumptions—namely, that transactions are processed on a First-Come, First-Served (FCFS) basis with Priority Fee Auctions (PFA). Moreover, most rollups (e.g., Arbitrum, Optimism, Base, Unichain, ZKsync) operate private mempools. MEV opportunities on L2s are thus limited to back-running strategies such as arbitrage and liquidations, or probabilistic (statistical) front-running or sandwiching~\cite{torres2024rolling}. Due to the low latency of rollups (200 ms - 2 s), the extraction of MEV becomes a latency race to be the first to submit transactions to the sequencer.

In March~2024, the Dencun upgrade~\cite{2024EthereumRoadmap} introduced \textit{blobs}~\cite{Eth-Danksharding}, a temporary data storage mechanism designed to enhance rollup scalability. This upgrade led to a significant reduction in gas fees on L2s,  bringing them below \$0.01, ~\cite{dashboard@L2Fees}, but it was accompanied by a noticeable increase in the transaction revert rate~\cite{dashboard@reverted}. While L2 revert rates oscillated below 5\% prior to the upgrade, they rose to over 10\% afterward—by contrast, the revert rate on Ethereum mainnet remains around 1\%.
These trends suggest the emergence of new MEV strategies on rollups that might include submitting multiple copies of the same arbitrage transaction to race for inclusion, splitting transactions into smaller parts, and probabilistically targeting specific positions within the block (e.g., beginning or end) to exploit sandwich or back-running opportunities, which we investigate in details in this work.

Looking forward, reducing the rate of reverted transactions is essential for L2s. Although blob transactions were initially cheap, their costs are expected to rise, potentially impacting rollup scalability. Notably, reverted transactions are stored on-chain and consume resources, despite not modifying the blockchain state. To address these concerns, some rollups are exploring mitigation techniques. For example, Unichain~\cite{Unichain2024} is developing a \textit{revert protection mechanism} and MEV tax~\cite{Robinson2024Priority}, whereas Arbitrum has introduced \textit{Time Boost}~\cite{mamageishvili2023buying}, a mechanism that allows sequencers to prioritize MEV opportunities ahead of other transactions. This paper contributes both empirically and theoretically to better design and evaluations of such mechanisms.

\subsubsection*{Related Work.} 

Zhu et al.~\cite{zhu2024revert} present a game-theoretic model of MEV auctions under revert protection, showing theocratically that revert protection increases auction revenue, market efficiency, and blockspace usage by incentivizing deterministic participation. Our work empirically validates its assumptions. 

The empirical study of MEV, especially on L2 blockchains, remains relatively underexplored, and involves studies on non-atomic arbitrage~\cite{gogol2024quantifying,gogol2024CrossRollupMEV,oz2025pandora} and cyclic one~\cite{torres2024rolling}.
Solmaz et al.~\cite{solmaz2025optimistic} analyze cyclic arbitrage bots on L2s, termed \emph{optimistic MEV}, and find that they consume disproportionately more gas than they contribute in fees. In contrast, we show that for the broader class of reverted transactions, the reverse holds: sequencers earn disproportionately more from fees than they expend in gas.

\textit{TimeBoost}, introduced by Arbitrum (April~2025), is a transaction ordering policy for rollup sequencers that incorporates both transaction timestamps and bids~\cite{mamageishvili2023buyingtime}. Simulation results suggest that it can yield returns comparable to or exceeding those of simple FCFS ordering for L2 MEV searchers~\cite{fritsch2025mev}. 
\textit{Flashblocks}, introduced on \emph{Base} (July~2025) and \emph{Unichain} (August~2025), provide early transaction \textit{pre-confirmation} every 200\,ms by splitting each block into ten mini-blocks. Their effects on transaction ordering have not yet been empirically assessed—a gap this work aims to fill.

\subsubsection*{Contributions.}
This paper makes three main contributions.  

\textbf{(i) Empirical.} We provide the first systematic evidence that reverts on Ethereum rollups are structured MEV outcomes rather than random waste. Over 80\% are swap attempts, concentrated in USDC–WETH pools, generated by a small number of identifiable bot archetypes. The persistence of these patterns across Arbitrum, Base, Optimism, Unichain, and ZKsync shows that revert-heavy equilibria are structural to fast-finality rollups, independent of sequencing design. 

\textbf{(ii) Market inefficiency.} We demonstrate empirically that priority fee auctions on rollups are inefficient in practice. Transaction placement exhibits systematic mis-ordering, round-number bidding, and duplication spam, which inflate base fees for users while allowing sequencers to capture disproportionate revenues from failed attempts relative to their gas consumption. Moreover, the introduction of Flashblocks on Base and Unichain reduced global fee-ordering efficiency, localizing competition to the first slot while leaving subsequent slots largely insensitive to priority fees. These results challenge the assumption that PFAs allocate blockspace efficiently under fast-finality conditions.  

\textbf{(iii) Theoretical.} We develop a formal model that rationalizes the observed strategies. We prove that trade-splitting and duplication strictly dominate single-shot execution under convex adversarial loss, explaining the prevalence of fragmented swaps and high revert rates. 

By linking empirical evidence with formal analysis, our work establishes reverts as equilibrium outcomes of rational MEV strategies and highlights the need for protocol-level reforms to sequencing and revert protection on rollups.  

%% file: sections/02Background.tex
\section{Background on Rollups}
\label{sec:background}

According to the blockchain scalability trilemma~\cite{delmonte2020decentralization}, a blockchain can prioritize at most two of the following three properties: decentralization, security, and scalability. Ethereum, the leading DeFi blockchain by TVL, prioritizes decentralization and security. This design choice has led to network congestion, high gas fees, and limited throughput—approximately 12 transactions per second (TPS). To address these limitations, scaling solutions have been developed at both Layer 1 (L1) and Layer 2 (L2). L1 scaling introduces new blockchains with alternative consensus mechanisms~\cite{Lashkari@2021Consensus}, sharding~\cite{Wang@2019Sharding}, and independent infrastructure. In contrast, L2 scaling executes intensive computations off-chain, posting compressed results to the underlying L1~\cite{sguanci2021layer2,gangwal2022survey}.

\subsubsection*{Rollups.}
\textit{Rollups}~\cite{Thibault@2022rollups} are non-custodial L2 solutions that function as independent blockchains: they execute transactions, produce blocks, and periodically submit compressed transaction data to the L1. This design allows rollups to inherit the security guarantees of the underlying L1—e.g., Ethereum’s staked ETH—making it infeasible to tamper with rollup data without compromising L1 security.

\textit{Optimistic rollups}~\cite{Kalodner@2018Arbitrum} assume transactions are valid unless challenged. This assumption simplifies implementation and accelerates compatibility with the Ethereum Virtual Machine (EVM), helping optimistic rollups become the first to attract DeFi adoption. However, their fraud-proof mechanism introduces a withdrawal delay, typically enforced through a 7-day challenge period.

\textit{ZK-rollups}~\cite{Chaliasos@2024AnalyzingRollups} use zero-knowledge proofs (ZKPs) to validate state transitions. After a proof is generated off-chain by a \emph{prover}, it is verified on-chain by a smart contract (\emph{verifier}). This architecture ensures rapid finality and enhances compression—only proofs, not raw transaction data, need to be posted to L1. The trade-off is increased computational overhead due to proof generation.

A \textit{sequencer}~\cite{chaliasos2024formalrollups} is a key component of rollup infrastructure, responsible for ordering transactions, forming blocks, and bundling them into batches submitted to L1. This improves gas efficiency relative to L1 execution. Despite cryptographic guarantees from optimistic and ZK rollups, sequencers are currently centralized. Major rollups like Arbitrum, Optimism, Base, Unichain, ZKsync, and StarkNet rely on centralized sequencers—and in the case of ZK-rollups, centralized provers as well. Efforts are underway to decentralize both~\cite{motepalli2023soksequencers,wang2024mechanismprover}.

\begin{figure}[!ht]
  \centering
  \begin{tabular}{cc}
    \includegraphics[width=0.45\textwidth]{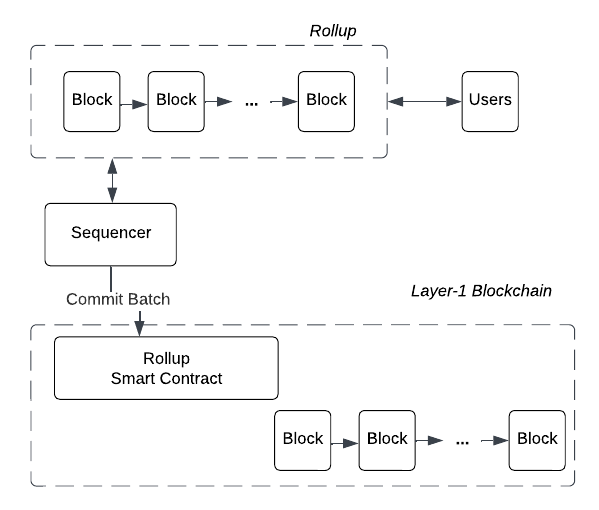} &
    \includegraphics[width=0.45\textwidth]{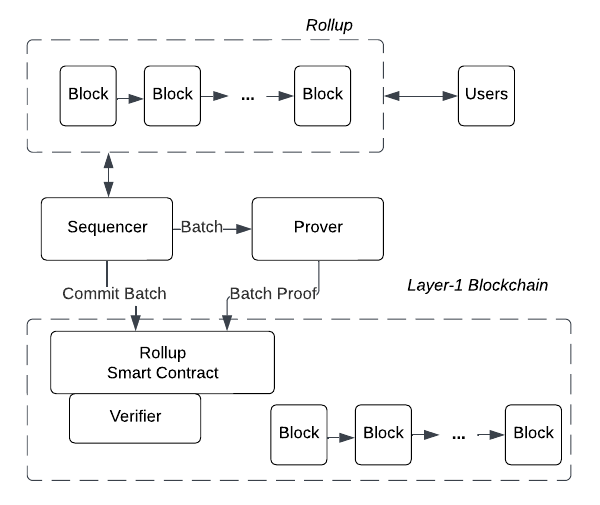} \\
    \small (a) Optimistic rollup & \small (b) ZK-rollup \\
  \end{tabular}
  \caption{Most rollups operate with centralized sequencers that maintain private mempools. These sequencers have exclusive control over transaction ordering within the rollup.
  }
  \label{fig:rollup}
\end{figure}

\textit{Blockchain finality} refers to the point at which a transaction becomes irreversible. On rollups, we distinguish between \textit{soft} and \textit{hard} finality. Soft finality occurs when a transaction is accepted by the sequencer and included in an L2 block, making it irreversible from the rollup's perspective. Hard finality is achieved once the corresponding batch is posted and confirmed on L1, securing the transaction with L1 guarantees. As rollups are currently centralized, users, bridges, and MEV searchers typically act upon soft finality, which occurs every 200ms–2s—much faster than Ethereum’s 12s L1 block time.\\



\noindent
\textbf{Gas Fees on L2s.}
Although reverted transactions do not alter the blockchain state, they are still included in the ledger and consume resources during execution.
In general, on Ethereum rollups, the total fee paid by a transaction comprises two main components:
\begin{itemize}
  \item \textit{L2 Execution Fee:} the cost of executing the transaction on the rollup's virtual machine.
  \item \textit{L1 Data Availability Fee:} the cost of posting transaction calldata to Ethereum L1, which ensures data availability and security.
\end{itemize}

\noindent
The L2 execution cost itself can be further decomposed into a \emph{base fee} and a \emph{priority fee}, leading to the following structure:

\begin{equation*}
\label{eq:cost}
\text{Gas Fee on L2} = 
\underset{\text{Gas Fee}}{
  \underbrace{
    \underset{\text{L2 Execution Fee}}{
      \underbrace{
        \text{Base Fee} + \text{Priority Fee}
      }
    } 
    + \text{L1 Data Fee}
  }
}
\end{equation*}

\noindent
These components can be computed from on-chain transaction data using the following formulas~\cite{buterin2021eip1559}:

\begin{align*}
\text{Execution Fee} &= \texttt{effective\_gas\_price} \times \texttt{gas\_used} \\
\text{Priority Fee}  &= \texttt{priority\_fee\_per\_gas} \times \texttt{gas\_used} \\
\text{Base Fee}      &= (\texttt{effective\_gas\_price} - \texttt{priority\_fee\_per\_gas}) \times \texttt{gas\_used}
\end{align*}

%% file: sections/03Methodology.tex
\section{Methodology and Data}
\label{sec:methodology}

We analyze successful and reverted transactions on five EVM-compatible rollups, Arbitrum, Optimism, Base, ZKsync, and Unichain, between January and July 2025, using Dune archive nodes as our data source. All analyzed rollups operate private mempools under centralized sequencers, with block times between 0.2s and 2s (vs.\ 12s on Ethereum). Table~\ref{tab:rollup_characteristics} summarizes their key parameters.

\begin{table*}[tb]
  \caption{Technical characteristics of analyzed Ethereum rollups: L2s rely on private mempools operated by centralized sequencers~\cite{oz2025pandora} with Priority Fee Auctions (PFAs). Arbitrum introduced TimeBoost (TB) sequencing on April 2025, Base introduced Flashblocks (FB) in July 2025, followed by Unichain in August 2025. The analyzed time period spans from January 2025 to August 2025. }
  \label{tab:rollup_characteristics}
  \centering
  \renewcommand{\arraystretch}{1.15} 
  \setlength{\tabcolsep}{6pt}        
  \small                             
  \begin{tabular}{lccp{1.5cm}p{1.4cm}c}
    \toprule
    \textbf{Rollup} & \textbf{Type} & \textbf{Mempool} & \textbf{Block Time} & \textbf{Order Flow} & \textbf{Launch} \\
    \midrule
    Arbitrum (ARB)  & Optimistic & Private & 0.25s & FCFS+TB & Aug 21 \\
    Base (BASE)     & Optimistic & Private & 2.0s . (0.20s FB)  & PFA       & Aug 23 \\
    Optimism (OP)   & Optimistic & Private & 2.0s  & PFA       & Dec 21 \\
    Unichain (UNI)  & Optimistic & Private & 2.0s . (0.20s FB) & PFA       & Apr 24 \\
    ZKsync (ZK)     & ZK         & Private & 1.0s  & FCFS       & Mar 23 \\
    \bottomrule
  \end{tabular}
\end{table*}

\subsubsection*{Data Collection.}
Users interact with blockchains through atomic transactions, which revert entirely if contract conditions are violated (e.g., slippage limits in swaps). Although reverted transactions do not emit event logs, their execution steps remain visible in the \textit{Traces} table. We collect transaction data from Dune’s \textit{Transactions} and \textit{Traces} tables, and supplement successful transactions with event data from \textit{EventLogs}.

\paragraph{Swap Labeling.}
A transaction is labeled as a swap if it interacts with a DEX pool or router and attempts token transfers. Labels from Dune’s \texttt{dex.trades} are extended manually to cover missing or newly deployed exchanges.

\paragraph{Atomic MEV Detection.}
We identify cyclic (atomic) MEV transactions following the methodology of Torres~\cite{torres2024rolling} and Flashbots~\cite{flashbotsAtomicMEV}.

\subsubsection*{Graph Construction for Reverted Swaps.}  
Reverted transactions are challenging to analyze because they do not emit application-layer event logs. The EVM, however, still records their internal execution steps as \emph{execution traces}. We leverage this property to construct a \emph{call graph} for each reverted transaction. Formally, the execution graph of a transaction~$\tau$ is a directed rooted tree $G_\tau = (V, E)$, where each node $v \in V$ corresponds to an internal execution step (i.e., a trace entry), and each edge $(v_i \rightarrow v_j) \in E$ denotes a call relationship initiated by $v_i$. Each node is annotated with its operation type (e.g., \texttt{CALL}, \texttt{DELEGATECALL}, \texttt{STATICCALL}), caller and callee addresses, ETH value transferred, calldata and returndata, and the execution status (success or revert).

To identify attempted DEX swaps, we traverse~$G_\tau$ and match \texttt{to} addresses against a canonical list of pools and routers (from Dune’s \texttt{dex.trades}, extended manually). For Uniswap v2/v3 (and forks), we directly verify pool or router calls with ERC-20 transfer selectors. In contrast, Uniswap v4 requires a different approach: because all liquidity is mediated through a shared \texttt{PoolManager} and hooks, we detect swaps by locating calls to the \texttt{PoolManager}, interactions with known tokens (e.g., WETH, USDC), and calldata patterns consistent with \texttt{modifyPosition}, \texttt{swap}, or hook-triggered functions. Transactions satisfying these conditions are labeled as \emph{reverted swaps}, and the corresponding subgraph is extracted for further analysis.

%% file: sections/05EmpiricalResults.tex
\section{Empirical Analysis of Reverted Transactions}
\label{sec:empirical}


\paragraph{Revert Rate Characteristics.}

\begin{figure*}[t]
    \begin{subfigure}[t]{\textwidth}
        \centering
        \includegraphics[width=\textwidth]{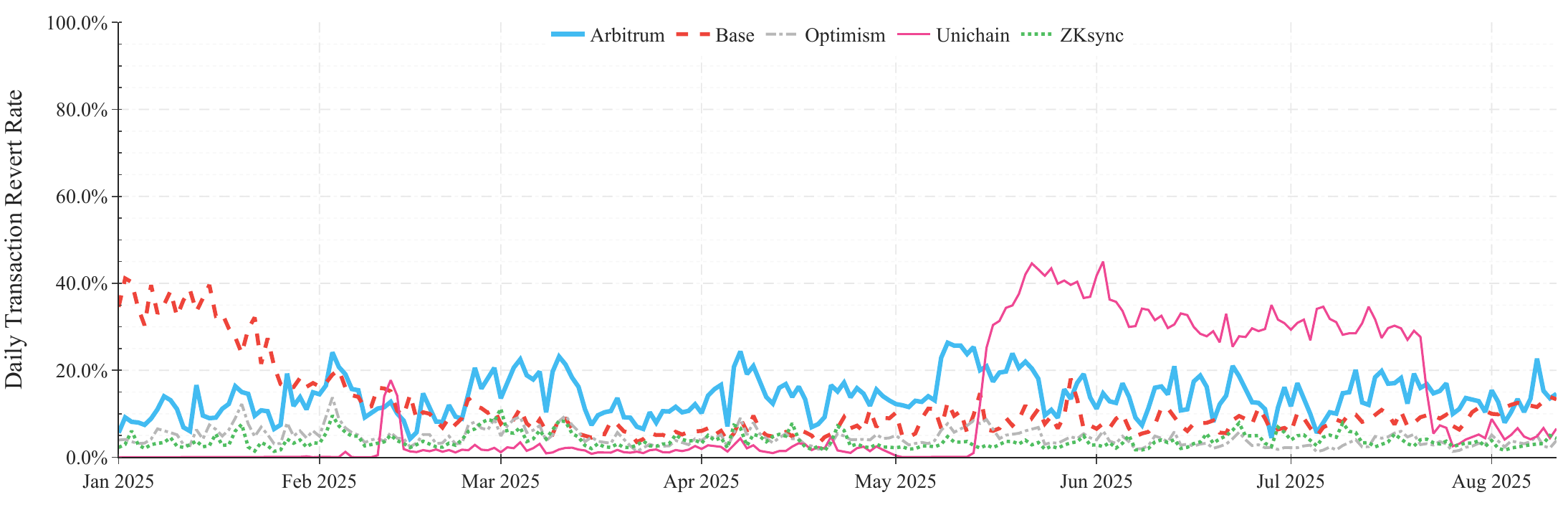}
        \caption{Daily transaction revert rates on Ethereum L2s.}
        \label{fig:revert-rate-all}
    \end{subfigure}

    \vspace{1em} 

    \begin{subfigure}[t]{0.48\textwidth}
        \centering
        \includegraphics[width=\textwidth]{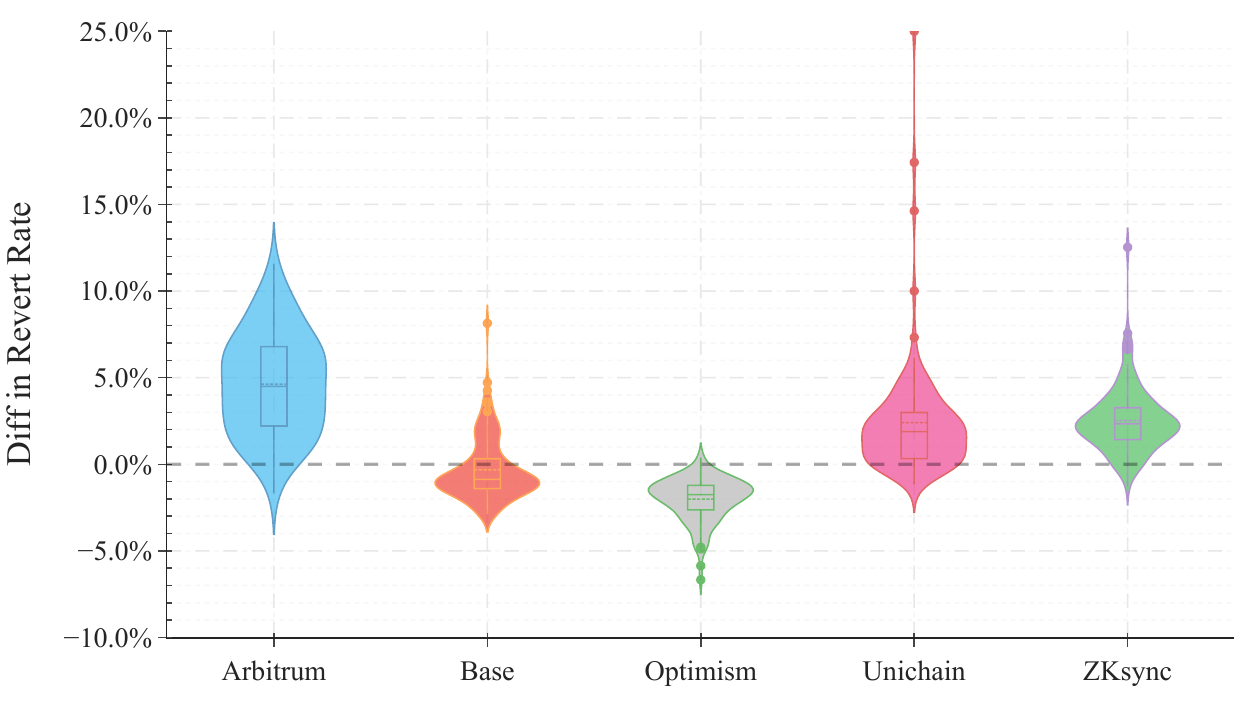}
        \caption{Difference in daily revert rates between transactions that set priority fee and all transactions.}
        \label{fig:diff-revert-rate}
    \end{subfigure}%
    \hfill
    \begin{subfigure}[t]{0.48\textwidth}
        \centering
        \includegraphics[width=\textwidth]{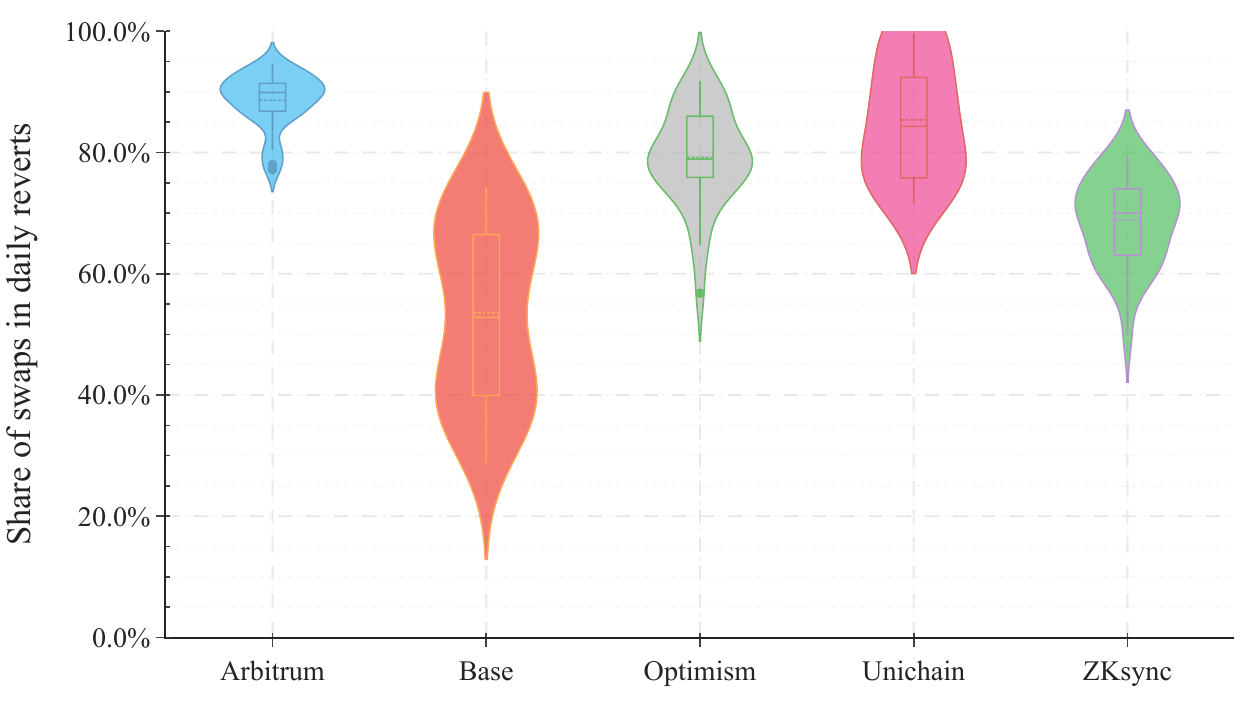}
        \caption{Daily share of swaps among reverted transactions.}    
        \label{fig:swaps-in-reverted-tx}
    \end{subfigure}
    
    \caption{Analysis of transaction reverts from January to August 2025. 
    (b) A positive difference on most L2s indicates that priority-fee transactions, often linked to MEV, revert more frequently. 
    (c) Reverts are predominantly swaps, reflecting MEV/DEX activity.}
    \label{fig:revert-rate-analysis}
\end{figure*}

Figure~\ref{fig:revert-rate-all} shows that daily revert rates ranged from 5\% to 40\% across L2s, with peaks up to 20\% on Arbitrum. On Unichain, revert rates increased sharply in mid-May and remained close to 40\% until July.
Figure~\ref{fig:swaps-in-reverted-tx} demonstrates that most of reverted transactions are swaps. In particular, \textbf{more than 80\% of reverted transactions are swaps} on Arbitrum, Optimism, and Unichain, suggesting their connection to arbitrage. Due to incomplete labeling, these values represent lower bounds, especially on Base.

\paragraph{Attributes of Reverted Swaps.}

\begin{figure}[t]
  \centering
    \includegraphics[width=1.1\textwidth]{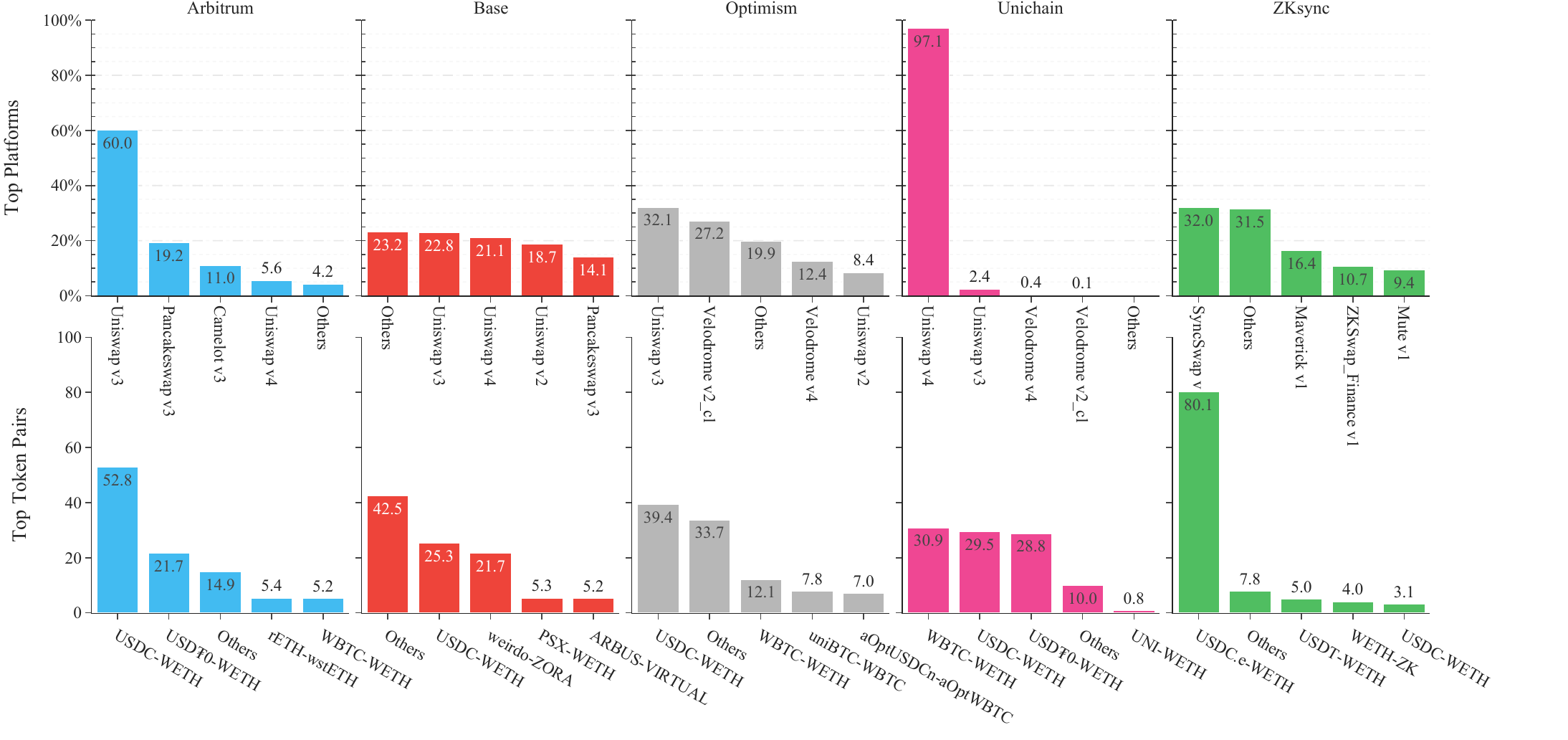} 
\caption{Breakdown of reverted swaps by target DEX and token pair. 
Most reverted swaps target WETH--USDC pools and Uniswap (v3).}
    \label{fig:reverted-swap-analysis}
\end{figure}

A detailed breakdown of reverted swaps characteristics is presented in Figure~\ref{fig:reverted-swap-analysis}. The \textbf{majority of reverted swaps interact with USDC--WETH pools and Uniswap v3}, suggesting that most originate from MEV searchers targeting highly liquid pairs. On Arbitrum, Uniswap v3 dominates with over 60\% of reverted swaps. On Unichain, Uniswap v4 is the primary target, accounting for more than 90\% of all reverted swaps.  Base exhibits a more balanced distribution across Uniswap v2, v3, and v4 (each $\sim$20\%). The most frequently targeted pair is USDC--WETH (over 50\% of reverts on Arbitrum),  with additional focus on USDC--USDT and WBTC--WETH.

\paragraph{Priority Fees and Auction Participation.}

Users may specify a priority fee per gas to incentivize inclusion. Figure~\ref{fig:priority-fees-per-gas}, in Appendix, shows the distribution of priority fees set by reverted transactions. Strikingly, nearly half of reverted transactions—and almost all on ZKsync—include a priority fee despite the absence of a priority fee auction on these L2s. On Unichain, nearly all reverted transactions set the priority fee to exactly 1 wei, indicating minimal auction engagement.  
Figure~\ref{fig:diff-revert-rate} compares revert rates of priority-fee transactions with all transactions. Base and Optimims exhibit a positive difference, indicating that \textbf{transactions setting priority fees revert more often}. This counterintuitive result is explained by their association with MEV strategies, which we evaluate further.

\paragraph{Block Positioning of Reverts.}

\begin{figure}[t]
  \centering
    \includegraphics[width=\textwidth]{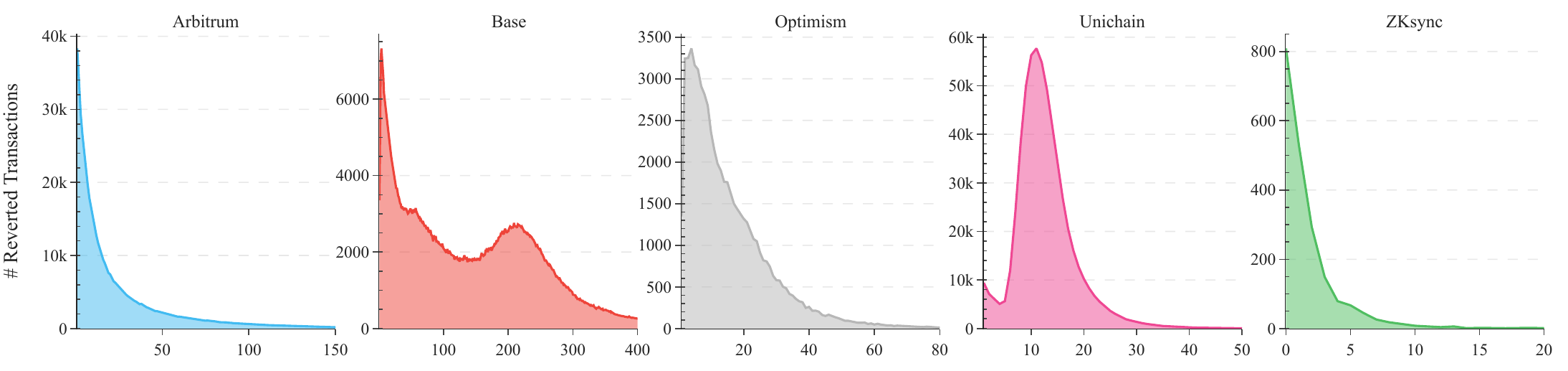} 
    \caption{Index of reverted transactions within blocks. 
    On Arbitrum and ZKsync, reverts cluster at the block start; 
    on Base, Optimism, and Unichain, they peak around the fourth position.}
    \label{fig:reverted-tx-index}
\end{figure}

Figure~\ref{fig:reverted-tx-index} analyzes intra-block positioning. On Arbitrum and ZKsync, reverted transactions cluster at the very beginning of blocks, consistent with intense low-latency races. On Base, Optimism, and Unichain, reverts peak later (around position 4), indicating different MEV execution strategies under longer block times.  

Figure~\ref{fig:index-vs-fee} in Appendix~\ref{app:figures} presents the relationship between the priority fee set and the block index position of reverted transactions across L2s. On Arbitrum and ZKsync, no clear pattern emerges (FCFS L2s). In contrast, on Base, Optimism and Unichain we observe a consistent trend: transactions offering higher priority fees are typically included closer to the beginning of the block. An additional notable observation is the concentration of reverted transactions around ``round-number'' priority fees (e.g., 100 wei, 1 million wei). This suggests a strong competition for inclusion at these discrete bidding levels and highlights the prevalence of suboptimal bidding strategies among MEV bots. 

Appendix~\ref{app:PFA} analyzes the impact of Flashblocks on the efficiency of PFAs. The results show that \textbf{only the first Flashblock exhibits efficient fee-based ordering}, while in subsequent slots the priority fee has negligible influence on transaction position within the block. This suggests that \textbf{priority fee auctions are not functioning as intended on fast-finality rollups}, where latency dominates over bidding.

\begin{figure}[t]
  \centering
  \includegraphics[width=1\textwidth]{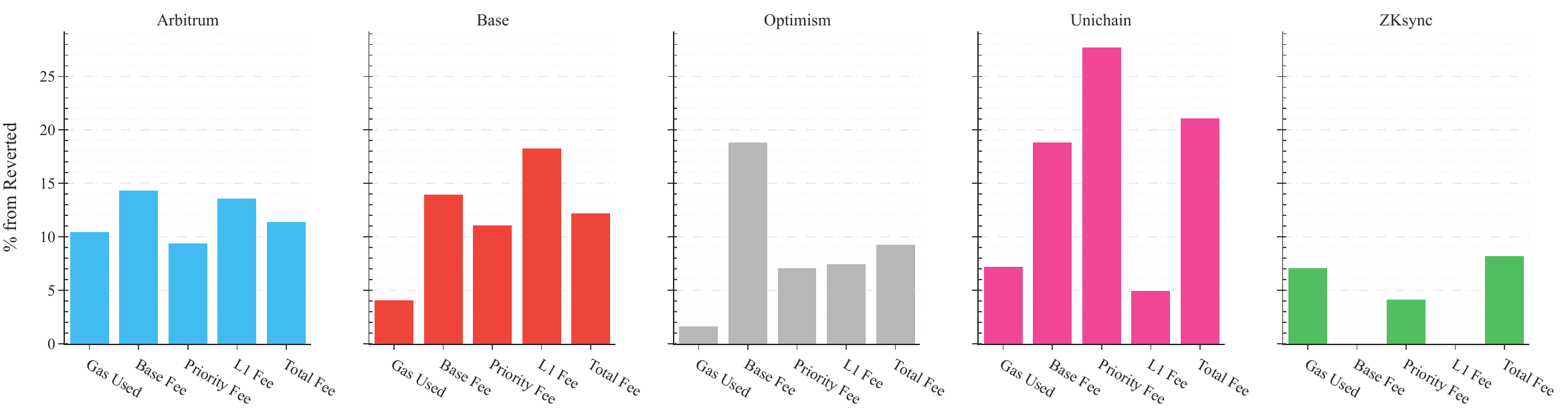}
  \caption{Share of reverted transactions in gas usage and L2 fee revenue. Reverted transactions contribute disproportionately more to fee revenue than to gas usage.}
  \label{fig:pct-gas-fees}
\end{figure}

\paragraph{Gas Usage and Fees.}
Figure~\ref{fig:pct-gas-fees} shows that reverted transactions \textbf{contribute disproportionately more to sequencer fee revenues than to gas usage}. Across all L2s, their share of total fees paid exceeds their share of gas consumed.  On Base and Unichain, reverted transactions generate $\sim$20–25\% of priority fee revenue. A longitudinal view of daily gas usage and fee contributions is provided in Appendix~\ref{app:figures} (Figure~\ref{fig:pct-daily-gas-fees}).

%% file: sections/06Clustering.tex
\section{Address Clustering and MEV Strategies}
    \label{sec:clustering}


We apply a clustering approach to identify MEV strategies and bots involved in reverted transactions. 
\textbf{We cluster blockchain addresses based on their occurrence as the recipient ("to" address) in reverted transactions}, since these addresses correspond to contracts that execute MEV strategies rather than EOAs that merely submit them. Unlike senders, which are often one-off spam shells, recipient contracts (e.g., routers, arbitrage bots, or swap helpers) are stable points of interaction that generate internal calls (traces) to DEX routers and liquidity pools, where reverts typically occur. Clustering at this level allows us to identify recurring patterns of MEV activity, distinguish user contracts from automated bots, and uncover characteristic patterns of failed or repeated arbitrage attempts.

On May 20th, 2025, we identified \num{759}, \num{1938}, \num{578}, \num{99}, \num{68} such addresses on Arbitrum, Base, Optimism, Unichain, and ZKsync, respectively. Due to the relatively small sample sizes on Unichain and ZKsync, we focus our clustering analysis on Arbitrum, Base, and Optimism.

For each one of those instances, we considered multiple factors, such as 
number of transactions, number of distinct addresses that called the analyzed address,
daily coverage (number of hours with at least one transaction),
revert rate,
and found the following set of features significant:
\begin{enumerate}
    \item \texttt{Swap \%} - percentage of swaps among the successful transactions,
    \item \texttt{Atomic \%} - percentage of atomic arbitrage swaps among the successful transactions,
    \item \texttt{Revert/Block} - average number of reverted transactions in a block among blocks with reverted transactions,
    \item \texttt{Success/Block} - average number of successful transactions in a block among blocks with successful transactions,
    \item \texttt{Revert No} - number of reverted transaction in a day.
\end{enumerate}

We perform Principal Component Analysis (PCA) on the standardized version of our selected features. The first five principal components explain approximately $\{26.5\%, 24.1\%, 18.6\%, 16.1\%, 14.7\%\}$ of the variance, respectively. Hence, the variance is distributed relatively evenly, and no single dominant component captures the majority of the variability in the data.

\begin{figure}[t]
  \centering
  \begin{subfigure}{.48\textwidth}
    \centering
    \includegraphics[width=0.95\linewidth]{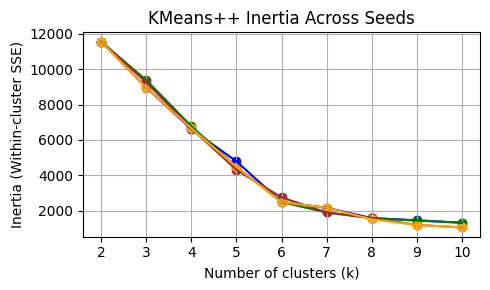}
    \caption{Inertia plot pointing to optimal clusters.}
    \label{fig:inertia-plot}
  \end{subfigure}%
  \hfill
  \begin{subfigure}{.48\textwidth}
    \centering
    \includegraphics[width=0.9\linewidth]{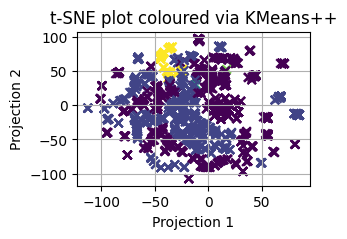}
    \caption{t-SNE projection of addresses.}
    \label{fig:tsne}
  \end{subfigure}

  \vspace{1em}
  \begin{adjustbox}{max width=\textwidth}
  \footnotesize
  \renewcommand{\arraystretch}{1.05}
  \setlength{\tabcolsep}{6pt}
  \begin{tabular}{c l r r r r r r r}
  \toprule
  Gr. & Name & \shortstack{\# of \\Recipients} & \shortstack{Swap\\\%} & \shortstack{Atomic\\\%}
  & \shortstack{Reverts\\/ Block} & \shortstack{Success\\/ Block} & \shortstack{Revert\\Rate} & \shortstack{Reverts\\/ Day} \\
  \midrule
  0 & Passive Receivers                & 1630 & 0.0\%   & 0.0\%   & 1.00   & 1.00   &  4.0\%  & 2 \\
  1 & Single-Shot Arbitrageurs         & 1032 & 100.0\% & 0.0\%   & 1.00   & 1.00   & 25.0\%  & 6 \\
  2 & Tx Relays / Spam Shells & 1    & 0.0\%   & 0.0\%   & 36.90  & 85.30  & 19.9\%  & 4{,}135 \\
  3 & Split-Trade Arbitrageurs         & 2    & 72.2\%  & 0.0\%   & 4.40   & 3.05   & 58.2\%  & 150{,}894 \\
  4 & End-of-Block Spammers            & 2    & 50.0\%  & 0.0\%   & 196.65 & 2.95   & 51.0\%  & 27{,}937 \\
  5 & Atomic Duplicators               & 68   & 100.0\% & 67.2\%  & 1.00   & 1.00   & 46.8\%  & 98 \\
  \bottomrule
  \end{tabular}
  \end{adjustbox}

  \caption{We cluster our data points relating to reverted transactions. 
  The inertia plot suggests an optimal clustering into~$6$ groups, 
  visualised via a t-SNE projection and summarised in the table.}
  \label{fig:clustering}
\end{figure}

We proceed to cluster our data via KMeans++ and experiment with different seeds to initiate the algorithm with, and with the possible number of target clusters to look for (we test for values ranging from~$2$ to~$10$). The related inertia plot is shown in Figure~\ref{fig:inertia-plot}, which clearly suggests $6$ as the optimal number of clusters to look for. We therefore save the KMeans++ clustering obtained for such~$6$ groups, and plot the t-SNE projection of our points in Figure~\ref{fig:tsne}, where the points are color-coded by the KMeans++ clustering labels. 

Figure~\ref{fig:clustering} summarizes the features for each cluster, in order to help with their interpretation and discussion. For each group, we report the number of points belonging to it, but also show the average value of each feature that was considered during our clustering phase. In general, we can make the following observations.

\paragraph{Group 0: Passive Receivers.}
This is the largest cluster, containing the majority of recipient contracts. 
These addresses exhibit almost no interaction with swaps (\texttt{swap\%} $\approx 0$), 
very low atomic arbitrage involvement, and only a small number of reverted transactions per block. 
Their activity is sporadic, with limited daily coverage, suggesting that they correspond to 
non-MEV infrastructure contracts or passive user contracts that occasionally receive reverted calls, 
rather than specialized arbitrage bots. In other words, Group~0 forms the \emph{background noise} 
of the reverted transaction landscape.

\noindent
\textit{Example.} 
The contract \href{https://basescan.org/address/0x42d80fdf14f13ca7019fd5a77fd53f1c1f4099a6}{\texttt{0x42d\ldots9a6}} on Base 
illustrates this passive-receiver pattern, with rare reverted calls but no consistent swap activity.

\paragraph{Group 1: Single-Shot Arbitrageurs.}
Addresses in this group almost exclusively execute swaps 
(\texttt{swap\%} close to~$1$), with on average one successful swap and one revert per block. 
The strategy is binary: either the arbitrage succeeds or the transaction reverts. 
This pattern is consistent with a \textbf{single-shot CEX--DEX arbitrage strategy}, 
where arbitrageurs attempt to capture price discrepancies between a centralized exchange 
and a single DEX pool with one atomic swap. When the opportunity vanishes during block inclusion, 
the trade simply fails.

\noindent
\textit{Example.} 
The address \href{https://optimistic.etherscan.io/address/0x436d53d45370b36c9b00740f1004e18182138795}{\texttt{0x436d\ldots795}} 
on Optimism repeatedly engages Uniswap v3 pools in this binary success-or-revert pattern.

\paragraph{Group 2: Transaction Relays / Spam Shells.}
This cluster contains only a few addresses, but with extremely high numbers of both 
reverted and successful transactions per block. 
They do not rely on swaps (\texttt{swap\%} $\approx 0$), 
indicating that these may be \emph{auxiliary contracts or router helpers} 
that aggregate or forward bundles of calls. 
Their high throughput per block suggests that they act as \textbf{transaction relays or 
spam shells}, rather than genuine arbitrage bots.

\noindent
\textit{Example.} 
The contract \href{https://basescan.org/address/0xc40d4aa7a84e48a0bb746f1c97d7023ba8055f90}{\texttt{0xc40d\ldots f90}} on Base 
shows this pattern of dense transaction forwarding without swap involvement.

\paragraph{Group 3: Split-Trade Arbitrageurs.}
This group produces the \emph{largest number of reverted transactions per day}, 
with revert rates exceeding~50\%. 
It contains only a handful of addresses, but each shows a distinct pattern: 
transactions are swaps, and if successful they appear multiple times within the same block, 
otherwise they revert. 
This behavior is consistent with a \textbf{CEX--DEX arbitrage strategy with trade splitting}, 
where a larger arbitrage opportunity is decomposed into several smaller swap attempts to 
mitigate slippage and latency risk. 
Such splitting maximizes expected returns but naturally generates many failed attempts, 
explaining the observed dominance in reverted transactions.

\noindent
\textit{Examples.} 
\href{https://basescan.org/address/0xb7b8f759e8bd293b91632100f53a45859832f463}{\texttt{0xb7b\ldots 463}} on Base 
targets Uniswap v3 USDC--WETH swaps exclusively, while 
\href{https://arbiscan.io/address/0xcb43d843f6cadf4f4844f3f57032468aadd9b95c}{\texttt{0xcb4\ldots 95c}} on Arbitrum 
splits trades across multiple pools (USDC--WETH, USDT--WETH, WBTC--WETH, USDC--ARB), 
both exemplifying this split-arbitrage style.

\paragraph{Group 4: End-of-Block Spammers.}
This cluster is characterized by an unusually high number of reverts per block 
combined with very few successful swaps. 
The addresses appear to send repeated transactions that fail more often than not, 
with activity peaking at end-of-block positions. 
This pattern fits a \textbf{spam-based, end-of-block sniping strategy}, 
in which bots flood the sequencer with revert-prone transactions in hopes that one 
lands in a favorable position to exploit short-lived price dislocations caused by noise traders. 

\noindent
\textit{Examples.} 
Two Base addresses illustrate this strategy: 
\href{https://basescan.org/address/0x7304133f4f94f2cf2d26359427fa0c03c65a3bfb}{\texttt{0x730\ldots bfb}} 
and 
\href{https://basescan.org/address/0x055a2f007145fe418680a8570a353be10a106298}{\texttt{0x055\ldots 298}}, 
both sending repeated spam-like swaps with high revert rates.

\paragraph{Group 5: Atomic Duplicators.}
Although relatively small in size, this cluster stands out with very high 
\texttt{swap\%} and \texttt{atomic\%} values, and a revert rate above~50\%. 
Addresses here engage almost exclusively in arbitrage via swaps, 
often across known DEX routers, and operate around the clock 
(\texttt{timezone coverage} $\approx 24$). 
The strategy resembles \textbf{atomic arbitrage with duplication}, 
where multiple parallel attempts of the same swap are fired within a block 
to maximize the chance of capturing a fleeting price gap. 
Compared to Group~1, these bots are more aggressive and maintain continuous coverage 
of opportunities across time zones.

\noindent
\textit{Example.} 
The Optimism address 
\href{https://optimistic.etherscan.io/address/0x44793fe8cc3c2bb881ba461694148f8fb49bd146}{\texttt{0x4479\ldots 146}} 
exemplifies this aggressive, around-the-clock atomic duplication behavior.

%% file: sections/07Theory-new.tex
\section{Theoretical Model of Trade Splitting and Duplication}
    \label{sec:model}

The prevalence of duplicated swaps and fragmented trade attempts observed in Section~\ref{sec:empirical} and~\ref{sec:clustering}  suggests that arbitrageurs strategically split transactions to optimize their chances of successful execution under slippage and latency risks. Our theoretical model captures these dynamics, providing a foundation for interpreting why such patterns of reverts and clustering emerge in practice. 
It abstracts away from priority fees. In particular, priority fees they are absent (e.g., Arbitrum, zkSync) or empirically negligible (e.g., Unichain), allowing us to study duplication and splitting without loss of generality.

\noindent
\textit{Token pair.}
There is a single token pair $(X,Y)$ traded on
\begin{itemize}
  \item a \textbf{constant-product AMM} with reserves $(x,y)$ and fee $f\in[0,1)$;
  \item a \textbf{centralized exchange (CEX)} that quotes a \emph{fixed} price $P_c>0$ expressed in $Y$ per $X$ and always settles.
\end{itemize}

\noindent
\textit{AMM swap mechanics.}
If an arbitrageur swaps $q>0$ units of $X$ with the AMM liquidity pool, she receives
\begin{equation}\label{eq:dy}
  \Delta y(q)
  =
  \frac{y(1-f)q}{x+(1-f)q},
\end{equation}
and the reserves in the AMM liquidity pool update to $(x+(1-f)q,\ y-\Delta y(q))$.
Formula~\eqref{eq:dy} is the standard CPMM payout.


\noindent
\textit{Costs.}
\begin{itemize}
  \item A fixed per-swap overhead $c_g\ge 0$ (gas \& latency).
  \item If the swap fails, the arbitrageur is left \emph{long} $q$ units of $X$;
        liquidating this inventory incurs a penalty $\phi\ge 0$ (possibly $0$).
\end{itemize}

\noindent
\textit{Arbitrage objective.}
The trader wishes to arbitrage a \emph{total} size $D>0$ (in token~$X$) by

\begin{enumerate}
  \item buying $D$ units of $X$ on the CEX for $P_cD$ units of $Y$;
  \item selling those units into the AMM, possibly split into $n\in\mathbb N$
        equal chunks of size $q=D/n$, or send multiple copies of the full trade.
\end{enumerate}

\noindent
\textit{Per–swap success.}
For one attempted swap of size $q$, we have the following notation.
\begin{itemize}
  \item\emph{Success} ($p(q)$): receive $\Delta y(q)$ from AMM, offset the $P_cq$ spent on CEX.
  \item\emph{Failure} ($1-p(q)$): CEX trade clears, AMM trade reverts; trader still paid $P_cq$ on CEX and may liquidate at cost $\phi$.
  \item\emph{Overhead} $c_g$ is incurred regardless.
\end{itemize}

\noindent
\textit{Adversarial loss.}
We model on-chain execution hazards (e.g., sandwiching) via an
\emph{adversarial loss} function $L:[0,\infty)\to\mathbb{R}_+$ applied to the submitted size $q$.
Economically, $L(q)$ captures expected value extracted by adversaries around the trade.
In CFMMs, curvature of the pricing curve together with user slippage bounds implies that the
attackable ``wedge'' scales \emph{more than linearly} with size, so it is natural to assume:
\begin{assumption}[Convex adversarial loss]\label{ass:L}
$L$ is nondecreasing, strictly convex, and $L(0)=0$.
\end{assumption}
This abstraction bundles sandwich/visibility costs and similar effects; constants and modeling details
enter only through the shape of $L(\cdot)$.

\noindent
\textit{Conditioning convention across subsections.}
There are two sensible ways to place $L$ relative to success probability $s$:
(i) \emph{conditional} $L$ (loss realized only if the swap executes), and
(ii) \emph{unconditional} $L$ (already averaged over success).
To avoid double counting while keeping formulas simple, we adopt the following convention:
\begin{itemize}
\item{Section~\ref{sec:splitting} (single-attempt splitting).}
Here the per-attempt success probability $s$ is treated as a \emph{constant}. We therefore
\emph{absorb $s$ into $L$} and, with slight abuse of notation, write $L(\cdot)$ for the
\emph{expected} adversarial loss per attempt (already scaled by $s$). This preserves convexity and
keeps statements compact.
\item {Section~\ref{sec:dup-with-L} (duplicate submissions).}
Here the success probability depends on the number of copies $K$ (namely $1-(1-s)^K$).
To prevent spurious scaling of loss with $K$, we keep $L(\cdot)$ \emph{conditional on success}
and multiply it by the overall success probability when computing expectations.
\end{itemize}


\subsection{Splitting trades}\label{sec:splitting}
We first show that under convex adversarial loss, arbitrageurs can strictly improve their expected profit by splitting a large trade into smaller slices.
\begin{proposition}[Splitting under convex adversarial loss]\label{prop:convex-splitting}
Fix a CFMM pool and a total input size $D>0$. Let the per-attempt success probability be constant $p(q)\equiv s\in(0,1]$.
Let $L:[0,\infty)\to\mathbb{R}_+$ satisfy Assumption~\ref{ass:L}.
Let the per-attempt fixed overhead
$\kappa = c_g + (1-s)\phi, \ (\kappa\ge 0).$
Denote by $\Delta y(q)$ the CFMM output of swapping size $q$ from the initial state, and by $\{\Delta y_i\}_{i=1}^n$ the \emph{incremental} outputs of executing $n$ equal slices $q=D/n$ back-to-back in the same pool without any external state refresh.
Define the expected profit from a single attempt of size $q$ as
\[
\pi(q) = s\Delta y(q) - P_cq - L(q) - \kappa ,
\]
and the total expected profit from splitting into $n$ equal slices as
\[
\Pi(n)=\sum_{i=1}^n \Big(s\Delta y_i - P_cD/n - L(D/n) - \kappa\Big).
\]
Then
\begin{equation}\label{eq:Pi-diff-general}
\Pi(n) - \Pi(1)
= s\!\left(\sum_{i=1}^n \Delta y_i - \Delta y(D)\right)
+ L(D) - nL(D/n)
- (n-1)\kappa .
\end{equation}
In particular, under same-pool, same-block execution, $\sum_{i=1}^n \Delta y_i = \Delta y(D)$ and
\begin{equation}\label{eq:Pi-diff-samepool}
\Pi(n) - \Pi(1) = \big(L(D) - nL(D/n)\big) - (n-1)\kappa.
\end{equation}
Hence splitting strictly improves expected profit (relative to one big trade) whenever
\[
L(D) - nL(D/n) > (n-1)\kappa .\footnote{Proof in Appendix~\ref{app:proofs}}
\]
\end{proposition}

\begin{remark}[Conditioning of the adversarial loss $L$]
In CFMM sandwiches the attacker’s profit is realized only if the victim’s swap executes. Since in this section the per-attempt success probability $s$ is treated as \emph{constant}, we absorb it into $L$ and interpret $L(\cdot)$ as the \emph{expected} adversarial loss per attempt (already scaled by $s$). This preserves convexity and leaves Proposition~\ref{prop:convex-splitting} unchanged (only coefficients rescale, e.g., in the quadratic case).
\end{remark}

\begin{corollary}[Quadratic adversarial loss]\label{cor:quadratic-L}
If $L(q)=\alpha q^2$ with $\alpha>0$, then
\[
\Pi(n) - \Pi(1) = \alpha D^2\!\left(1 - \frac{1}{n}\right) - (n-1)\kappa.
\]
Treating $n$ as a positive real and differentiating,
\[
\frac{d}{dn}\big[\Pi(n) - \Pi(1)\big]
= \alpha D^2\!\left(\frac{1}{n^2}\right) - \kappa.
\]
Setting this to zero gives the continuous optimum $n^\star_{\mathrm{cont}} = D\sqrt{\alpha/\kappa}$. The optimal integer choice is
\[
n^\star = \max\Big\{1,\ \big\lfloor n^\star_{\mathrm{cont}}\big\rceil\Big\},
\]
i.e., the nearest integer.
\end{corollary}

\subsection{Duplicate-Submission Spam}\label{sec:dup-with-L}
Fix a slice size $q>0$ and let $s:=p(q)\in[0,1]$ be the success probability of a single submission of size $q$.
If a trader submits \(K\in\mathbb{N}\) identical copies such that at most the earliest one can execute, then the success event occurs iff at least one copy lands. Under the standard independent-arrival approximation,  
the success probability is
\[
s_K(q)=1-\bigl(1-s\bigr)^K.
\]

We model adversarial extraction (e.g., sandwiching) by a loss function $L(q)$ (satisfying Assumption~\ref{ass:L}) that is \emph{conditional on success}: if one copy executes, the realized gross benefit $\Delta y(q)$ is reduced by $L(q)$, and if no copy executes, no such adversarial loss is realized.\footnote{In Section~\ref{sec:splitting} we treated $s$ as a constant \emph{per attempt} and therefore absorbed it into $L$ without loss of generality. In duplication, the success probability depends on $K$ via $1-(1-s)^K$, so keeping $L$ \emph{conditional on success} avoids artificially scaling adversarial loss with $K$. See Remark~\ref{rem:conditional-L-dup} below.}

Let $W:=\Delta y(q)+\phi$. The expected utility of choosing $K$ copies is
\begin{align}\label{eq:U-K-with-L}
\begin{split}
U(K;q)&=\underbrace{\big[1-(1-s)^K\big]\big(\Delta y(q)+\phi - L(q)\big)}_{\text{success term}}
-P_c q-\phi-Kc_g
\\ &=\underbrace{\big[1-(1-s)^K\big]\big(W-L(q)\big)-Kc_g}_{\displaystyle F_L(K;q)}-P_c q-\phi.
\end{split}
\end{align}
The term $P_c q+\phi$ does not affect the optimizer in $K$, so we maximize $F_L(K;q)$.

We next establish conditions under which duplicating the same trade multiple times is optimal, and derive the formula for the optimal number of copies.
\begin{proposition}[Optimal duplication]\label{prop:duplication-L}
Fix $q>0$ and define $A:=W-L(q)=\Delta y(q)+\phi-L(q)$. Then:
\begin{enumerate}
\item[(i)] \textbf{No duplication.} If $A\le 0$ or $sA\le c_g$, it is optimal to send a single copy: $K^\star(q)=1$.
\item[(ii)] \textbf{Optimal duplication.} If $A>0$ and $sA>c_g$, the optimal number of copies is the largest integer $K$ such that
\[
F_L(K;q)-F_L(K-1;q)=(1-s)^{K-1}\,sA-c_g\ge 0.
\]
Equivalently,
\begin{equation}\label{eq:Kstar-with-L}
K^\star(q)
=
1+\left\lfloor \frac{\ln\!\big(c_g/(sA)\big)}{\ln(1-s)} \right\rfloor\footnote{Proof in Appendix~\ref{app:proofs}.}
\end{equation}
\end{enumerate}
\end{proposition}



\begin{remark}[Why $L$ is conditional here]\label{rem:conditional-L-dup}
Only the earliest-included copy can execute; all other copies revert. Adversarial extraction (e.g., sandwich attacks) requires the victim trade to execute. Hence $L$ is naturally specified {conditional on success} in the duplication problem, and it is multiplied by the total success probability $1-(1-s)^K$ in Equation~\eqref{eq:U-K-with-L}. By contrast, in Section~\ref{sec:splitting} we analyzed a single attempt with {constant} success probability $s$; there, absorbing $s$ into $L$ is without loss of generality and simplifies the exposition. Keeping $L$ conditional here prevents scaling of adversarial loss with the number of duplicate submissions.
\end{remark}


%% file: sections/08Discussion.tex
\section{Discussion}
\label{sec:discussion}

\paragraph{Reverts as Equilibrium.}  
Our results establish that reverted transactions on rollups are not accidental noise but equilibrium outcomes of rational MEV strategies. Splitting and duplication strictly dominate single-shot execution under convex adversarial loss, and empirically we observe exactly these patterns across L2s. High revert rates are therefore an endogenous response to the interaction of fast block times, private mempools, and fee rules, rather than evidence of inefficient or malicious actors alone. Interpreting reverts as equilibrium behavior reframes them from wasted blockspace to a predictable by-product of current sequencing designs, thus reverts are best understood as equilibrium outcomes shaped by sequencer policy, not anomalies.

\paragraph{Sequencer Mechanisms: Flashblocks and TimeBoost.}  
The introduction of Flashblocks on Base and Unichain, and TimeBoost on Arbitrum, illustrates how small changes in sequencer policy reshape rather than eliminate MEV equilibria (Appendxi\ref{app:PFA}). Flashblocks concentrate fee competition in Slot~1, where revert probability rises sharply, while subsequent slots remain insensitive to fees. TimeBoost introduces fee sensitivity at the block head, but FCFS ordering persists elsewhere. In both cases, revert-heavy equilibria shift to new focal points rather than disappear. These designs redistribute congestion costs but do not alleviate the overall welfare loss for users.

\paragraph{Implications for Revert Protection.}  
Reverted transactions contribute disproportionately to sequencer revenues—up to one-quarter of all priority fees on some rollups—despite consuming far less gas. This asymmetry raises the question of whether reverted transactions should be exempt from base and L1 data fees, while still paying priority fees to compensate sequencers for processing. Such a design could reduce welfare losses for ordinary users, but care is required: transactions that currently settle without altering state could be strategically adapted to qualify as reverts, creating new avenues for exploitation. The challenge is thus to design revert protection that lowers systemic waste while preventing gaming by sophisticated bots. Hybrid fee schemes that penalize duplication without punishing isolated reverts may be a promising direction.

\paragraph{Cross-Rollup Generality.}  
The persistence of inefficiencies across Arbitrum, Base, Optimism, Unichain, and ZKsync suggests that revert-heavy equilibria are an inherent risk of fast-finality rollups, rather than an artifact of a particular sequencing mechanism.

%% file: sections/09Conclusions.tex
\section{Conclusions}
\label{sec:conclusions}

This paper examined the economics of reverted transactions on Ethereum rollups through a combination of empirical measurement, clustering, and theoretical modeling. We showed that reverts are not random noise but systematic outcomes of MEV strategies: more than 80\% are swap attempts, heavily concentrated in USDC–WETH pools, and generated by identifiable bot archetypes such as split-trade arbitrageurs, atomic duplicators, and end-of-block spammers. Our analysis further revealed that priority fee auctions on rollups fail to allocate blockspace efficiently. Mis-ordering, round-number bidding, and duplication spam inflate base fees for users while enabling sequencers to capture disproportionate revenues from failed attempts relative to their gas consumption. 

To rationalize these dynamics, we developed a formal model proving that trade-splitting and duplication strictly dominate single-shot execution under convex adversarial loss, thereby explaining the prevalence of fragmented swaps and high revert rates. Together, these findings reframe reverts as equilibrium features of current rollup fee markets rather than wasteful anomalies.  

The implications are clear: without protocol-level reforms to sequencing, revert protection, and transaction fee mechanisms, duplication-heavy equilibria will persist, leaving rollup fee markets both inefficient and welfare-distorting. 

%% file: sections/99SequncerOrdering.tex
\section{Impact of Flashblocks and TimeBoost on Transaction Ordering by Sequencer}
\label{app:PFA}

Flashblocks, a form of early transaction \textit{pre-conformation}, were introduced on \emph{Base} in July~2025 and on \emph{Unichain} in August~2025. 
Both L2s, developed with the OP~Stack operate with block times of approximately $2$ seconds, and Flashblocks offer the pre-confiramtion every 200mn, 

Arbitrum sequencer blocks are produced at an interval of $\sim$200--250ms, providing substantially faster inclusion latency than OP~Stack chains prior to the introduction of Flashblocks. However, Arbitrum sequencer does not consider priority fee, when ordering the transaction (FCFS rule). Thus, in April~2025, Arbitrum introduced \emph{TimeBoost}, a modification of the sequencer ordering policy that establishes a priority lane for the boosted transactions.

\subsection{Mechanism Overview}

\paragraph{Flashblocks.} 
Flashblocks subdivide each L2 block into $N$ smaller units, typically $N=10$, referred to as \emph{preconfirmation slots}. 
Transactions are scheduled into these slots sequentially, providing users with a low-latency guarantee (preconfirmation) of inclusion before the full block is sealed. 
The design intention is twofold: (i) to reduce latency by offering preconfirmations every $\sim$200ms rather than once per block, and (ii) to create predictable inclusion windows. 

\paragraph{TimeBoost (Arbitrum).} 
TimeBoost augments the sequencing infrastructure with two components: a sealed-bid second-price auction and a dedicated \emph{express lane}. 
Valid transactions submitted to the express lane are sequenced immediately with no delay, while all other transactions experience a nominal deferment (default: 200ms). The auction winner is granted exclusive rights to control the express lane for predefined, temporary intervals, enabling fee-based competition for ultra-low-latency inclusion. The default block time for Arbitrum remains at $\sim$250ms, but under TimeBoost some transactions outside the express lane may be postponed to the subsequent block.

\begin{table}[t]
\centering
\caption{Comparison of Flashblocks and TimeBoost as Sequencer Ordering Mechanisms.}
\label{tab:flashblocks-timeboost}
\renewcommand{\arraystretch}{1.2}
\begin{tabular}{p{3cm} p{5cm} p{6cm}}
\toprule
\textbf{Feature} & \textbf{Flashblocks} & \textbf{TimeBoost} \\
\midrule
\textbf{Launch} & 
Base: July 2025 \newline 
Unichain: August 2025 & 
Arbitrum: April 2025 \\
\midrule
\textbf{Underlying stack} & 
OP Stack (Base, Unichain) & 
Arbitrum Nitro stack \\
\midrule
\textbf{Block time (baseline)} & 
$\sim$2s before Flashblocks & 
$\sim$200--250ms \\
\midrule
\textbf{Structure} & 
Block subdivided into $N$ \emph{preconfirmation slots} (typically $N=10$), each with $\sim$200ms latency guarantees. &
Single block structure maintained; introduces a new \emph{express lane} at the sequencer. \\
\midrule
\textbf{Auction type} & 
Implicit priority fee competition within each slot. &
Sealed-bid second-price auction determines control of the express lane. \\
\midrule
\textbf{Inclusion rule} & 
Transactions placed sequentially into slots; higher priority fees compete for earlier position in slots &
Winning bidder’s transactions in the express lane are sequenced immediately; others incur nominal delay ($\sim$200ms) and may be deferred to the next block. \\
\bottomrule
\end{tabular}
\end{table}

\subsection{Methodology}
Transaction ordering is normalized to the unit interval $[0,1]$, where $0$ denotes the block head and $1$ the block tail. 
On Flashblocks-enabled chains, each block is heuristically divided into ten equal slots to approximate sub-block ordering. 
Priority fees are grouped into logarithmically spaced bins, with an additional bin reserved for zero-fee transactions. 
For each fee bin and slot combination, we compute the median block position and the share of reverted transactions, and estimate regressions of median position on the log of the priority fee.

\begin{figure}[!ht]
  \centering
  \includegraphics[width=1.2\textwidth]{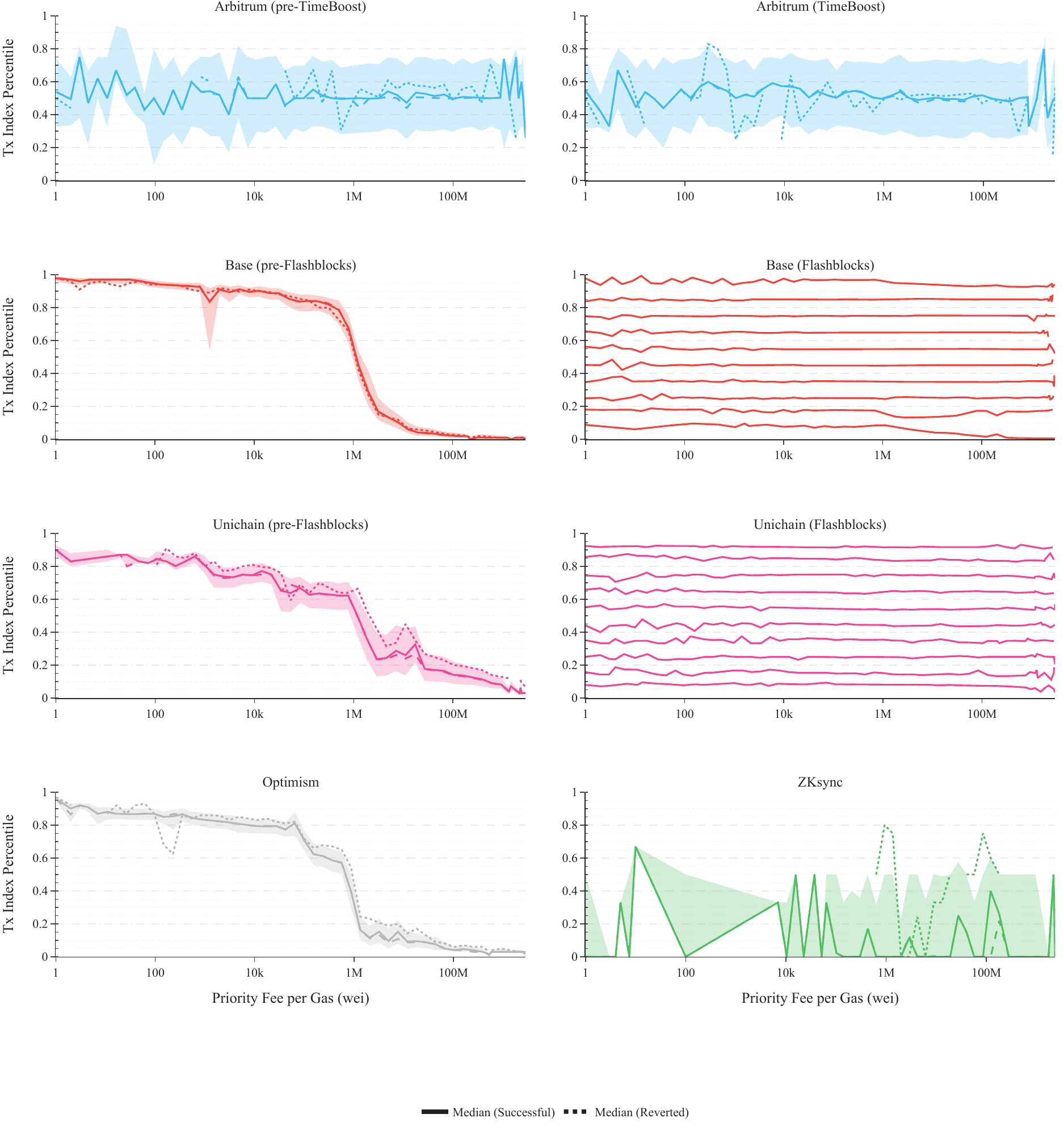}
  \caption{Median priority fee across transaction index percentiles, with interquartile range (IQR) bands, for each analyzed L2 chain. On Base, Optimism, and Unichain, higher priority fees are associated with earlier inclusion in the block (lower index positions), whereas no such pattern is observed on Arbitrum and ZKsync.}
  \label{fig:index-vs-fee}
\end{figure}

\subsection{Flashblocks: Results on Base and Unichain}

\paragraph{Before vs.\ After Heatmaps.}
Heatmaps of revert share ($\%$ of transactions reverted) across fee levels and slot position reveal a structural break. Prior to Flashblocks, revert clustering shows no discernible slot pattern, with mis-ordering dispersed throughout the block. After introduction, reverts concentrate at the block head (Slot~1), consistent with an auction-like mechanism.

\paragraph{Zero vs.\ Positive-Fee Comparison.}
Line plots of revert probability conditioned on fee regime show that zero-fee transactions remain stable across both periods, with consistently low revert shares. In contrast, positive-fee transactions are disproportionately exposed to reverts after Flashblocks, particularly in Slot~1.

\paragraph{Median Position vs.\ Fee (Slot-by-Slot).}
Slot-level median positions against fee reveal heterogeneous effects. Slot~1 exhibits a steep negative slope: higher fees reliably push transactions to the block head, but with elevated revert risk. Slots~2–9 display nearly flat relationships, indicating that fee has negligible ordering impact. Slot~10 shows a weaker but persistent secondary effect: transactions “tailgate” behind Slot~1 and inherit some revert exposure.

\paragraph{Regression Analysis.}
Slot-specific regressions of transaction position on log priority fee (Table~\ref{tab:slot_regressions_base_unichain}) confirm the descriptive patterns. 
For Base, Slot~1 exhibits the steepest negative slope, consistent with a functioning auction at the block head, while Slots~2--9 coefficients are near zero. 
Slot~10 shows a modest slope, consistent with secondary clustering. 
Unichain displays a similar structure: significant fee sensitivity in Slots~1, 6, 7, and 9, with other slots flat. 
These results indicate that Flashblocks localizes fee competition to a subset of slots rather than enforcing global ordering efficiency across the block.

\subsection{TimeBoost: Results on Arbitrum}

In April~2024, Arbitrum introduced \emph{TimeBoost}, a sequencing mechanism that prioritizes transactions submitted with valid priority bids. We apply the same empirical methodology as for Flashblocks: (i) heatmaps of revert share across fee levels and slots, (ii) zero vs.\ positive-fee comparisons, (iii) median transaction position vs.\ fee by slot, and (iv) regression analysis of fee–position sensitivity.

\paragraph{Before vs.\ After Heatmaps.}
Prior to TimeBoost, revert clustering exhibited no strong slot structure, with transactions dispersed across the block irrespective of fee. After TimeBoost, concentration at early positions (Slot~1) becomes evident, indicating auction-like prioritization.

\paragraph{Zero vs.\ Positive-Fee Comparison.}
Zero-fee transactions remain largely unaffected by TimeBoost, showing consistently low revert rates. By contrast, positive-fee transactions display increased fee sensitivity post-TimeBoost: higher fees are more strongly associated with early placement but also with elevated revert probability at the block head.

\paragraph{Median Position vs.\ Fee (Slot-by-Slot).}
Slot-level analysis reveals that prior to TimeBoost, the median position is only weakly responsive to fees. After TimeBoost, Slot~1 shows a sharp negative slope (higher fees reliably move transactions to the head), while Slots~2–9 remain flat, and Slot~10 displays a weaker tailgating effect, similar to Base and Unichain under Flashblocks.

\paragraph{Regression Analysis.}
Pooled regressions of transaction position on log priority fee (Table~\ref{tab:regressions_l2s}) show no statistically significant relationship either before or after TimeBoost. 
This confirms that Arbitrum’s sequencer continues to operate under a FCFS rather than a PFA for majority of transactions. 
While TimeBoost introduces an auction-like mechanism in principle, its effects are not observable in fee–position sensitivity at the block level, in sharp contrast to the localized ordering observed under Flashblocks.

\subsection{Discussion and Implications}

\paragraph{Flashblocks.}
Across Base and Unichain, the introduction of Flashblocks reshapes fee–ordering dynamics. The system approximates an auction only in Slot~1, while Slots~2–9 show no fee discrimination. Reverts disproportionately cluster at the block head, magnifying costs for participants bidding for early inclusion. Zero-fee transactions, by contrast, are stable and rarely reverted, highlighting the asymmetric burden of Flashblocks on fee-paying users.

\paragraph{TimeBoost.}
TimeBoost introduces meaningful fee sensitivity at the front of the block, aligning Arbitrum’s sequencing closer to a priority auction. However, ordering efficiency is limited: only Slot~1 reflects strong auction dynamics, while the majority of slots remain insensitive to fee bids. Reverts cluster more heavily at the block head post-TimeBoost, indicating that competitive bidding for early inclusion amplifies revert-based MEV strategies rather than eliminating them.

\begin{table}[htbp]
\centering
\scriptsize
\begin{tabular}{llrrrrr}
\toprule
Chain     & Slot & $\beta$ (log fee) & Std. Err. & t-stat & p-value & $R^2$ \\
\midrule
Base (Flashblocks)     & 1  & -0.005 & 0.001 &  -9.424 & $<0.001$ & 0.791 \\
                       & 2  & -0.001 & 0.000 &  -3.558 & $<0.001$ & 0.186 \\
                       & 3  &  0.000 & 0.000 &   0.938 & 0.353    & 0.028 \\
                       & 4  & -0.000 & 0.000 &  -1.272 & 0.210    & 0.064 \\
                       & 5  &  0.000 & 0.000 &   0.627 & 0.534    & 0.017 \\
                       & 6  & -0.000 & 0.000 &  -0.904 & 0.371    & 0.035 \\
                       & 7  & -0.000 & 0.000 &  -0.602 & 0.550    & 0.015 \\
                       & 8  &  0.000 & 0.000 &   0.856 & 0.396    & 0.028 \\
                       & 9  &  0.000 & 0.000 &   1.777 & 0.081    & 0.100 \\
                       & 10 & -0.002 & 0.000 &  -5.672 & $<0.001$ & 0.434 \\
\midrule
Unichain (Flashblocks) & 1  & -0.001 & 0.000 &  -4.163 & $<0.001$ & 0.366 \\
                       & 2  & -0.001 & 0.000 &  -1.986 & 0.052    & 0.114 \\
                       & 3  &  0.000 & 0.000 &   0.379 & 0.706    & 0.005 \\
                       & 4  & -0.000 & 0.000 &  -0.144 & 0.886    & 0.000 \\
                       & 5  &  0.000 & 0.000 &   0.521 & 0.605    & 0.008 \\
                       & 6  & -0.001 & 0.000 &  -3.050 & 0.004    & 0.206 \\
                       & 7  & -0.001 & 0.000 &  -3.157 & 0.003    & 0.217 \\
                       & 8  & -0.000 & 0.000 &  -0.393 & 0.696    & 0.005 \\
                       & 9  & -0.001 & 0.000 &  -4.507 & $<0.001$ & 0.418 \\
                       & 10 & -0.000 & 0.000 &  -1.931 & 0.060    & 0.110 \\
\bottomrule
\end{tabular}
\caption{Slot-level regressions of median transaction position on $\log(\text{priority fee})$ for Base and Unichain under Flashblocks. 
Strong fee-ordering persists in some slots (e.g., Base slots 1 and 10; Unichain slots 1, 6, 7, 9), while others show weak or no significant relationship. This reflects how Flashblocks localizes priority fee auctions to specific slots, reducing global fee-ordering efficiency.}
\label{tab:slot_regressions_base_unichain}
\end{table}

\begin{table}[htbp]
\centering
\small
\begin{tabular}{lrrrrr}
\toprule
Chain & $\beta$ (log fee) & Std. Err. & t-stat & p-value & $R^2$ \\
\midrule
Arbitrum (pre-TimeBoost)   & -0.001 & 0.002 &  -0.367 & 0.715 & 0.004 \\
Arbitrum (TimeBoost)       & -0.000 & 0.002 &  -0.153 & 0.879 & 0.001 \\
Base (pre-Flashblocks)     & -0.058 & 0.003 & -22.053 & $<0.001$ & 0.853 \\
Unichain (pre-Flashblocks) & -0.047 & 0.002 & -23.219 & $<0.001$ & 0.919 \\
Optimism                   & -0.053 & 0.002 & -25.961 & $<0.001$ & 0.882 \\
ZKsync                     & -0.004 & 0.006 &  -0.618 & 0.540 & 0.014 \\
\bottomrule
\end{tabular}
\caption{Pooled OLS regressions of median transaction position on $\log(\text{priority fee})$ across L2s. 
Negative coefficients indicate that higher fees lead to earlier block positions. Results show efficient fee-based ordering on Base, Unichain, and Optimism, while Arbitrum and ZKsync exhibit no significant relationship. This is consistent with their FCFS (first-come, first-served) sequencer policy rather than a priority fee auction (PFA).}
\label{tab:regressions_l2s}
\end{table}

%% file: sections/98Appendix.tex
\section{Proofs}\label{app:proofs}

\subsection{Proof of Proposition~\ref{prop:convex-splitting}}

To prove this proposition, we first need two lemmas. 

\begin{lemma}\label{lem:path-additivity}
Under same-pool, same-block execution with no external state refresh, the total output of $n$ equal slices sums to the output of one big trade:
\[
\sum_{i=1}^n \Delta y_i = \Delta y(D).
\]
\end{lemma}

\begin{proof}
Consider a constant-product AMM with reserves $(x,y)$ and invariant $k=xy$.
A single trade’s payout is $\Delta y(q)=y-\frac{k}{x+(1-f)q}$, consistent with \eqref{eq:dy}.
A single trade of (pre-fee) size $D$ therefore yields
$\Delta y(D) = y - \frac{k}{x + (1-f)D}$. Let $d := \frac{(1-f)D}{n}$ denote the \emph{post-fee} input per slice.
After $i-1$ slices the $x$-reserve is $x+(i-1)d$ and the $y$-reserve is $k/(x+(i-1)d)$.
The $i$-th slice outputs
$\Delta y_i = \frac{k}{x+(i-1)d} - \frac{k}{x+i d}$.
Summing telescopes,
\begin{align*}
\sum_{i=1}^n \Delta y_i
&= \left(y-\frac{k}{x+d}\right) + \left(\frac{k}{x+d}-\frac{k}{x+2d}\right) + \cdots + \left(\frac{k}{x+(n-1)d}-\frac{k}{x+nd}\right)\\
&= y - \frac{k}{x+nd}
= y - \frac{k}{x + (1-f)D} = \Delta y(D).
\end{align*} \qed
\end{proof}

\begin{remark}
Lemma~\ref{lem:path-additivity} assumes the $n$ slices are executed atomically, so the pool follows one deterministic path. This is a standard simplification to isolate the {frictionless} AMM mechanics. Any sandwiching or visibility costs are modeled separately by the convex loss term, not inside the mechanical output.

If interleaving is possible (public mempool), the proof still goes through with a small tweak: decompose the outcome of splitting into (i) a {path} effect and (ii) an {adversarial-loss} effect. 
\end{remark}

\begin{lemma}\label{lem:convexity-dividend}
If $L$ is strictly convex with $L(0)=0$, then for any $D>0$ and integer $n > 1$,
\[
L(D) - n\,L(D/n) > 0.
\]
\end{lemma}

\begin{proof}
Let $\lambda = 1/n \in (0,1)$ and $x=D$. By strict convexity of $L$ and $L(0)=0$,
$L\!\big(\lambda x + (1-\lambda) \cdot 0\big)
< \lambda L(x) + (1-\lambda)L(0)
= \frac{1}{n} L(D)$.
Since $\lambda x + (1-\lambda)\cdot 0 = \lambda D = D/n$, this is
$L(D/n) < \frac{1}{n} L(D)$.
Multiplying by $n$ yields $n\,L(D/n) < L(D)$, i.e., $L(D) - n L(D/n) > 0$.
\qed
\end{proof}

Now we can proceed to prove the proposition.
\paragraph{Proof of Proposition~\ref{prop:convex-splitting}}
Compute the difference directly:
\begin{align*}
\Pi(n) - \Pi(1)
&= \left[\sum_{i=1}^n \big(s\Delta y_i - P_cD/n - L(D/n) - \kappa\big)\right]
   - \big(s\Delta y(D) - P_c D - L(D) - \kappa\big) \\
&= s\!\left(\sum_{i=1}^n \Delta y_i - \Delta y(D)\right) 
   + \big(L(D) - nL(D/n)\big)
   - (n-1)\kappa,
\end{align*}
which is \eqref{eq:Pi-diff-general}. Under same-pool, same-block execution, Lemma~\ref{lem:path-additivity} yields $\sum_{i=1}^n \Delta y_i = \Delta y(D)$, so the first bracket vanishes and we obtain
\[
\Pi(n) - \Pi(1) = \big(L(D) - nL(D/n)\big) - (n-1)\kappa,
\]
as in \eqref{eq:Pi-diff-samepool}. By Lemma~\ref{lem:convexity-dividend}, $L(D) - nL(D/n)>0$ for $n>1$, hence $\Pi(n)>\Pi(1)$ whenever $L(D) - nL(D/n) > (n-1)\kappa$. \qed

\subsection{Proof of Proposition~\ref{prop:duplication-L}}
\begin{proof}
Write $g(K):=1-(1-s)^K$. Then $F_L(K;q)=g(K)\,A-Kc_g$. The discrete marginal gain from adding one more copy is
\[
\Delta_K F_L = F_L(K;q)-F_L(K-1;q) = \big(g(K)-g(K-1)\big)A - c_g = (1-s)^{K-1}\,sA - c_g.
\]
(i) If $A\le 0$, then $F_L(K;q)\le -Kc_g$ and $K^\star=1$. If $A>0$ but $sA\le c_g$, then $\Delta_1 F_L=sA-c_g\le 0$ and the objective decreases for all $K\ge 2$, hence $K^\star=1$.

(ii) If $A>0$ and $sA>c_g$, then $\Delta_1 F_L>0$ and $\Delta_K F_L$ is strictly decreasing in $K$ because $(1-s)^{K-1}$ decreases in $K$. Thus $F_L$ is strictly increasing until the smallest $K$ for which $\Delta_K F_L<0$, and strictly decreasing thereafter. The optimal $K^\star$ is the largest integer with $\Delta_K F_L\ge 0$,
which rearranges to \eqref{eq:Kstar-with-L}. \qed
\end{proof}

%% file: sections/99Appendix.tex
\newpage
\section{Additional Figures}
\label{app:figures}

\begin{figure}[tbh]
  \centering
    \includegraphics[width=\textwidth]{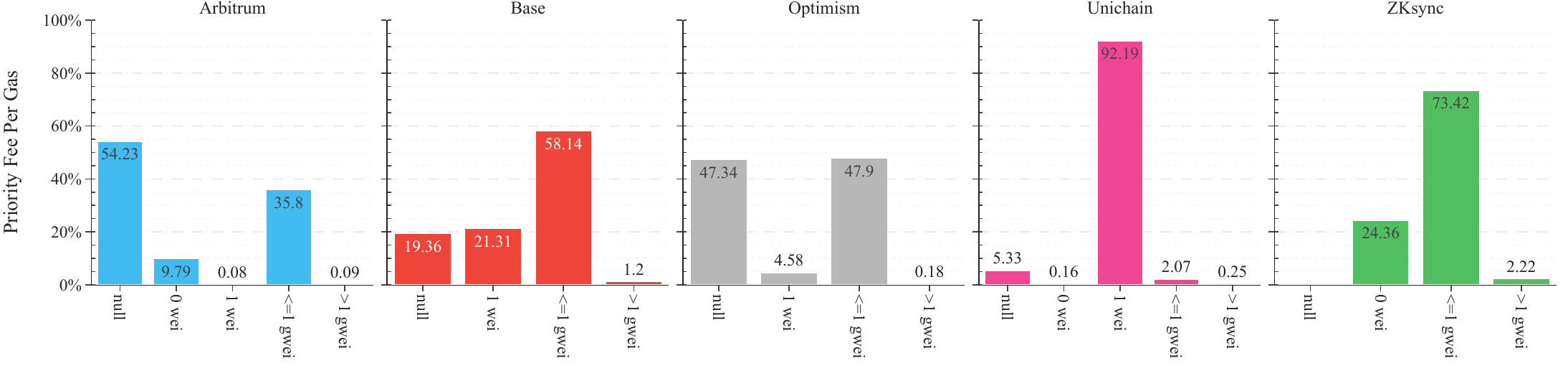} 
\caption{Priority fee per gas in reverted transactions across L2s. Base, Optimism, and Unichain implement PFAs, whereas Arbitrum and ZKsync rely on FCFS ordering. 
Strikingly, nearly half of reverted transactions—and almost all on ZKsync—include a priority fee despite the absence of a priority fee auction at these L2s.
}
    \label{fig:priority-fees-per-gas}
\end{figure}

\begin{figure}[!ht]
  \centering
  \includegraphics[width=1.2\textwidth]{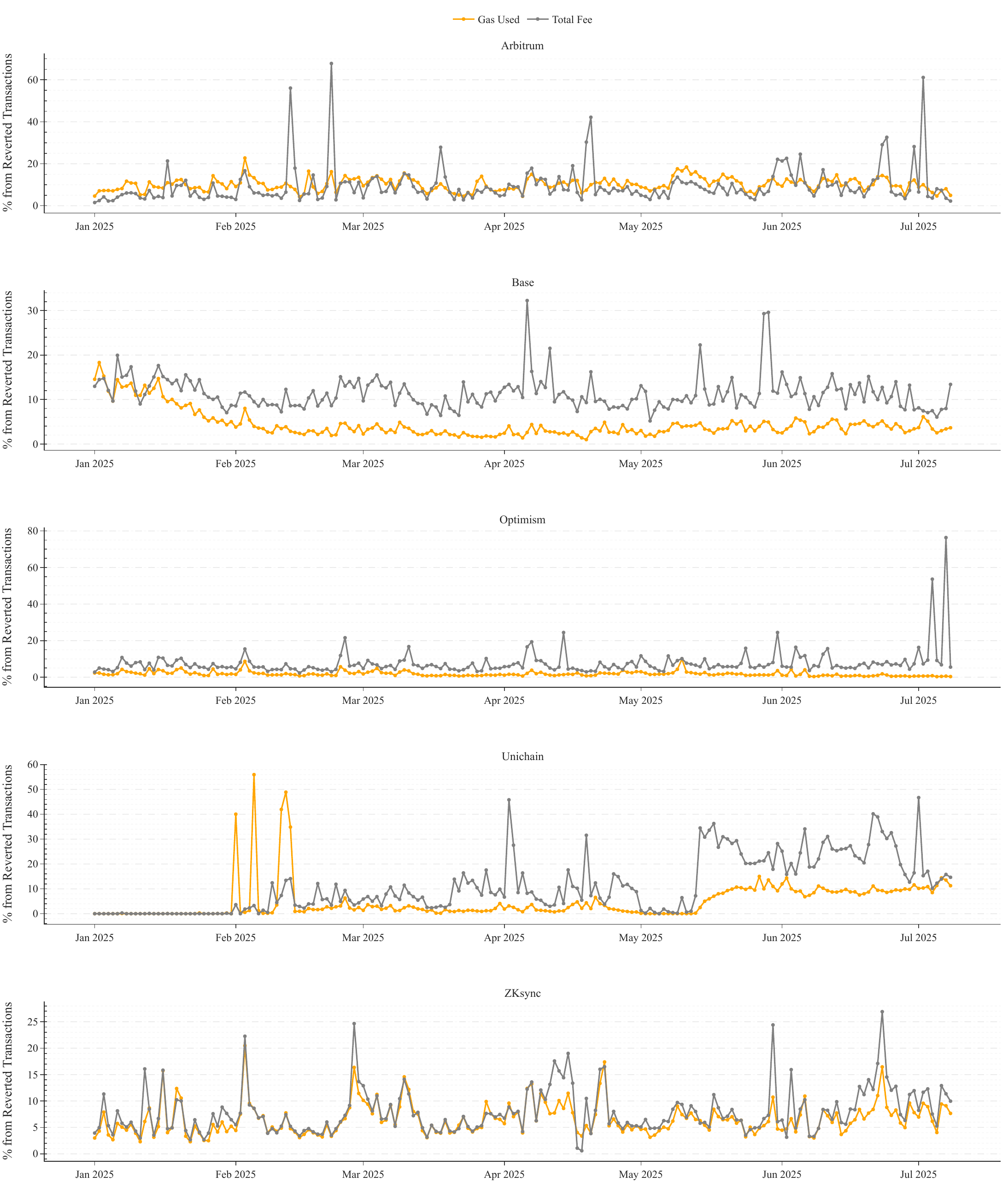}
  \caption{Daily share of reverted transactions in total gas consumption and fee revenue on each L2. Reverted transactions contribute disproportionately more to L2 fee revenues than to gas usage}
  \label{fig:pct-daily-gas-fees}
\end{figure}

\begin{figure}[!ht]
  \centering
  \includegraphics[width=1.2\textwidth]{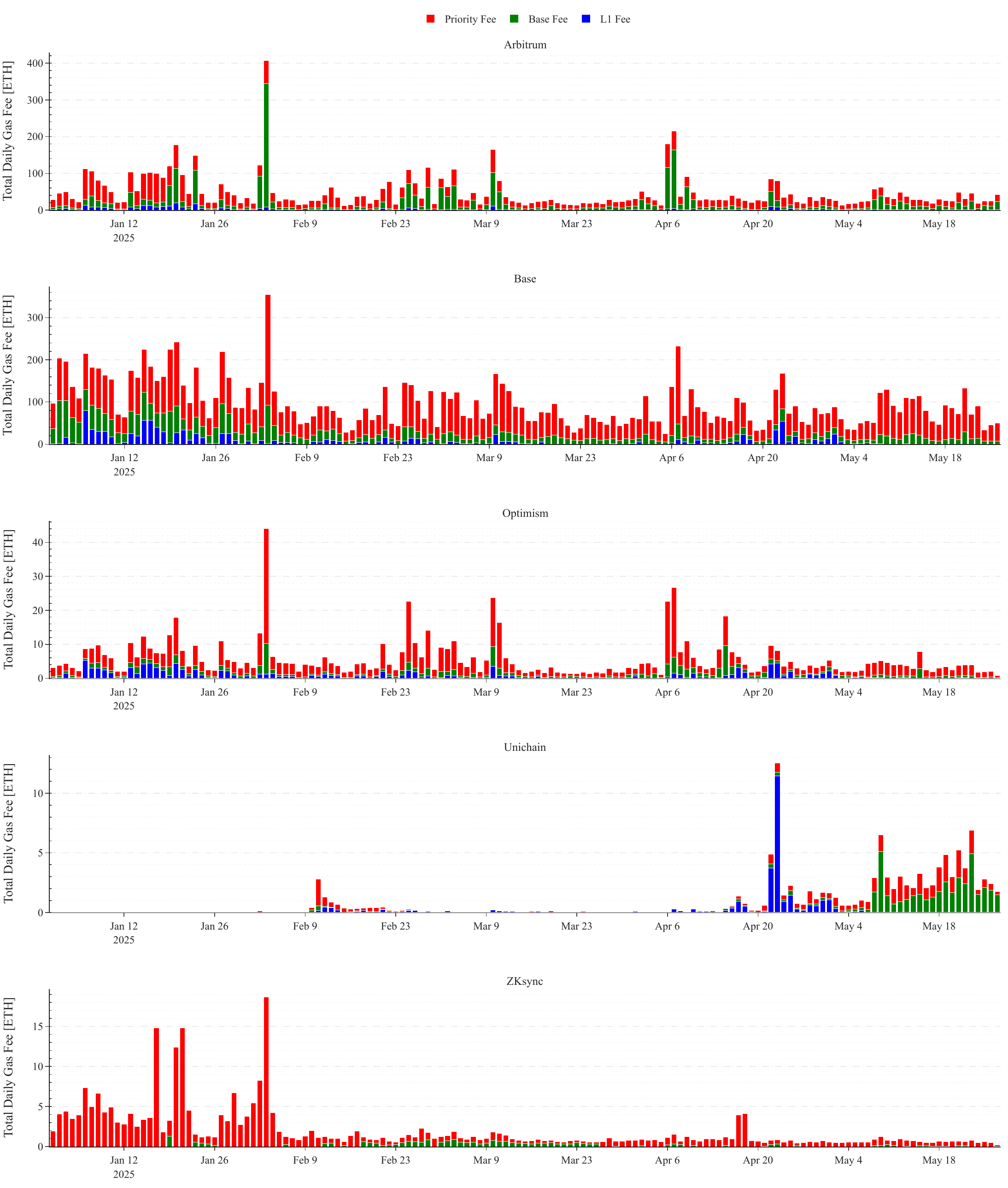}
  \caption{Daily total gas fees incurred by reverted transactions. The breakdown shows the L2 execution fee (split into base and priority components) and the L1 fee for calldata posting to Ethereum.}
  \label{fig:reverted_fees}
\end{figure}

\begin{figure}[t]
    \centering
    \includegraphics[width=1.2\linewidth]{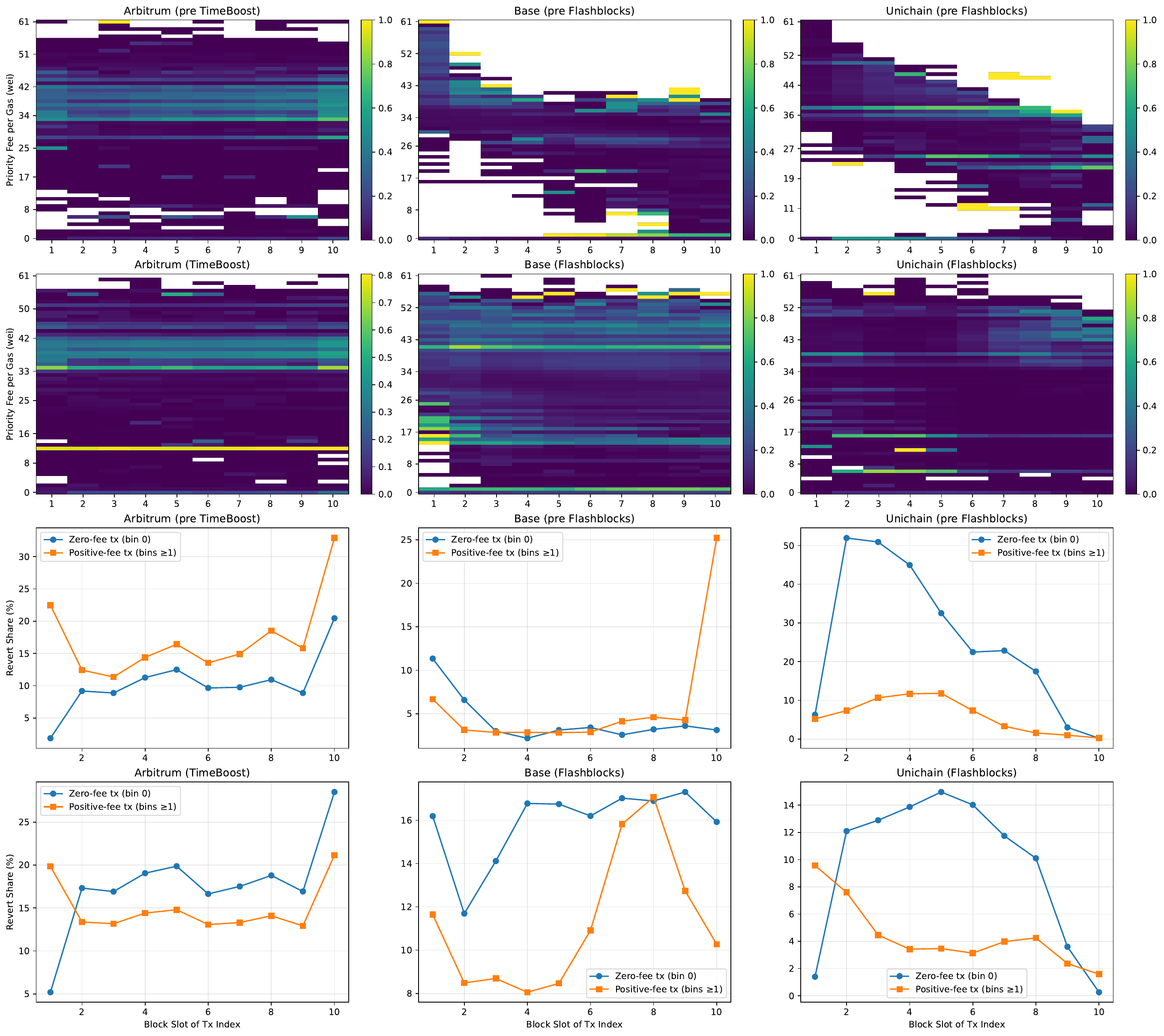}
    \caption{%
        Heatmaps of revert share ($y$-axis, color intensity) across log priority fee levels ($x$-axis) and Flashblock/TimeBoost slots (vertical axis), 
        with slot-level median position vs.\ fee lines overlaid. 
        Panels show before vs.\ after sequencing changes on Base, Unichain (Flashblocks), and Arbitrum (TimeBoost). 
        The heatmaps reveal localized auction behavior at Slot~1 after the mechanisms are introduced, while Slots~2–9 remain insensitive to fees. 
        Overlaid line plots highlight slot-by-slot fee responsiveness, confirming that only the block head exhibits auction-like dynamics.%
    }
    \label{fig:heatmaps-slotlines}
\end{figure}